\documentclass[preprint,proceedings]{rmaa}
\usepackage{rmaacite}

\usepackage{paralist}

\usepackage{psfrag,color}

\newcommand{\hatn}{{\mathbf {\hat n}}}

\SetYear{2002}
\SetConfTitle{Galaxy Evolution: Theory and Observations}

\title{Inflation at the Edges}

\author{M. Kamionkowski\altaffilmark{1}}
\altaffiltext{1}{California Institute of Technology, Pasadena, CA, USA}

\shortauthor{Kamionkowski}
\shorttitle{Inflation at the Edge}

\fulladdresses{\item California Institute of Technology, Mail Code
130-33, Pasadena, CA~~91125, USA (kamion@tapir.caltech.edu).}
\listofauthors{M. Kamionkowski}
\indexauthor{Kamionkowski, M.}

\abstract{Recent results from cosmic microwave background (CMB)
experiments verify several of the predictions of inflation,
while ruling out a number of alternative structure-formation
scenarios.  Given the successes of the theory, the obvious next
step is to press ahead and test inflation to the edge of all our
current and forthcoming observational abilities.  According to
the inflationary paradigm, galaxies and their large-scale
distribution in the Universe are remnants of inflation and can
thus be studied to learn more about inflation in the same way
that experimental particle physicists study the remnants of
high-energy collisions.  Here I discuss how studies of
galactic substructure, galaxies, clusters, large-scale
structure, and the CMB, may be used to learn more about
inflation.}
\resumen{Resultados recientes de experimentos del fondo de microondas
c\'osmico (FMC) prueban varias de las predicciones de la inflaci\'on
y descartan muchos de los escenarios de formaci\'on c\'osmica
alternativos.  Dado el \'exito de la teor\'{\i}a, el siguiente paso
obvio  es probar la inflaci\'on al filo de nuestras capacidades
observacionales actuales y por venir. De acuerdo al paradigma
inflacionario, las galaxies y su distribuci\'on a gran escala son
remanentes de la inflaci\'on por lo que pueden ser estudiadas para
explorarla m\'as a fondo, de la misma manera que un f\'{\i}sico de
 part\'{\i}culas experimental estudia los remanentes de colisiones
de alta energ\'{\i}a. A continuaci\'on har\'e un an\'alisis de
c\'omo estudios de subestructura gal\'actica, galaxias, c\'umulos,
estructura a gran escala y el FMC pueden ser usados para aprender
m\'as sobre la inflaci\'on.}

\addkeyword{Cosmology}

\begin{document}
\maketitle

\section{Introduction}
\label{sec:intro}

Inflationary cosmology \cite{Gut81,Lin82a,AlbSte82} has in
recent years had a number of
dramatic successes.  The inflationary predictions of a flat
Universe and nearly scale-invariant primordial density
perturbations with Gaussian initial conditions
\cite{GutPi82,Haw82,Lin82b,Sta82,BarSteTur83} have been found
to be consistent with a series of increasingly precise cosmic
microwave background (CMB) experiments
\cite{Miletal99,deBetal00,Hanetal00,Haletal02,Masetal02}.  Theorists
discuss
open-Universe and alternative structure-formation models,
such as topological defects, with {\it far} less frequency than
they did just three years ago.

Historically, when experimental breakthroughs confirm a
particular theoretical paradigm and eliminate others, progress
can be made at the edges---i.e., precision tests of the new standard
model.  In the case of inflation, a number of important questions should
be addressed.  For example, what is the physics responsible for
inflation?  What is the energy scale of inflation?  
In particular, we really do not understand why the
simple slow-roll model of inflation---really no more than a toy
model---works so well.  Might deviations from the simplest model
expected in realistic theories lead to small deviations from the
canonical predictions of inflation?  For example, is the density
of the Universe precisely equal to the critical density?  Are
there deviations from scale invariance on small distance scales
that arise as a consequence of the end of inflation?  Might
there be some small admixture of entropy perturbations
in addition to the predominant adiabatic perturbations?  Are
there small deviations from Gaussian initial conditions?

A variety of forthcoming CMB experiments will test the flatness
of the Universe with additional precision and determine the
primordial spectrum of perturbations with increasing accuracy.
CMB experiments and galaxy surveys and weak-lensing maps that
determine the mass distribution in the Universe today will test
Gaussian initial conditions.  Our understanding of galactic
substructure may shed light on the end of inflation.
Experimentalists are beginning to contemplate programs to
detect the unique polarization signature due to inflationary
gravitational waves.

Here, I briefly review several new probes of possible relics of
inflation; namely inflationary gravitational waves,
non-Gaussianity, and galactic substructure.

\section{Gravitational Waves and CMB Polarization}

\begin{figure}[!t]
  \includegraphics[width=\columnwidth]{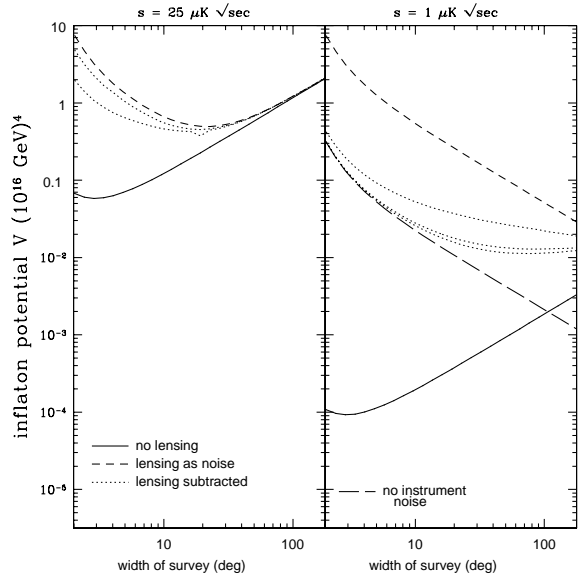}
  \caption{Minimum inflaton potential observable at
     $1\sigma$ as a function of survey width for a one-year
     experiment.  The left panel shows an experiment
     with NET $s=25\, \mu{\rm
     K}~\sqrt{\rm sec}$.
     The solid curve shows results assuming no CS
     while the dashed curve shows results including the effects
     of an unsubtracted CS; we take $\theta_{\rm
     FWHM}=5'$ in these two cases.  The dotted curves
     assume the CS is subtracted with $\theta_{\rm
     FWHM}=10'$ (upper curve) and $5'$ (lower curve).  Since
     the dotted curves are close to the dashed curve, it
     shows that these higher-order correlations will not be significantly
     useful in reconstructing the primordial curl for an
     experiment similar to Planck's sensitivity and resolution.  The
     right panel shows results for hypothetical improved
     experiments.  The dotted curves shows results with CS subtracted and 
     assuming $s=1\, \mu{\rm
     K}~\sqrt{\rm sec}$, $\theta_{\rm FWHM}=5'$, $2'$, and $1'$
     (from top to bottom). 
     The solid curve assumes $\theta_{\rm
     FWHM}=1'$ and $s=1\, \mu{\rm K}~\sqrt{\rm sec}$, and
     no CS, while the dashed curve treats CS as an additional
     noise.  The long-dash curve assumes  CS subtraction
     with no instrumental noise ($s=0$).  From \protect\scite{KesCooKam02}.}
  \label{fig:limits}
\end{figure}

\begin{figure}[!t]
  \includegraphics[width=\columnwidth]{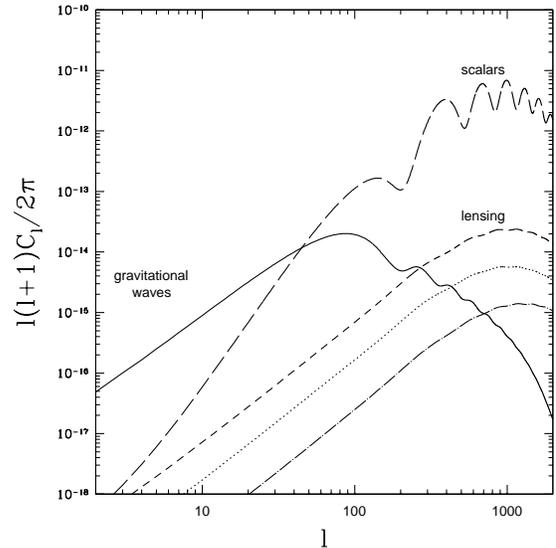}
  \caption{
     CMB polarization power spectra.  The
     long-dashed curve shows the dominant polarization signal in the
     gradient component due to scalar (density) perturbations. The solid
     line shows the maximum allowed curl polarization signal from the
     gravitational-wave background, which will be smaller if
     the inflationary energy scale is smaller than the maximum
     value allowed by COBE of $3.5 \times 10^{16}$ GeV.  
	The dashed curve shows the power
     spectrum of the curl component of the polarization due to
     CS.  The dotted curve is the CS contribution to the curl
     component that comes from structures out to a redshift of
     1; this is the level at which low-redshift lensing surveys
     can be used to separate the CS-induced polarization from
     the IGW signal. The dot-dashed line is the residual when
     lensing is separated with a no-noise
     experiment and 80\% sky coverage.  From
     \protect\scite{KesCooKam02}.}
  \label{fig:powerspectrum}
\end{figure}

One the most intriguing avenues toward further tests of
inflation is the gravitational-wave background.  In addition to
predicting a flat Universe with adiabatic perturbations,
inflation also predicts the existence of a stochastic
gravitational-wave background with a nearly-scale-invariant
spectrum \cite{AbbWis84}.   The amplitude of this inflationary
gravitational-wave background (IGW) is fixed entirely by the
vacuum-energy density during inflation, which is proportional to
the fourth power of the energy scale $E_{\rm infl}$ of the new
physics responsible for inflation.

Gravitational waves, like primordial density perturbations,
produce linear polarization in the CMB.  However, the
polarization patterns from the two differ.  
This can be quantified with a harmonic decomposition of the
polarization field.  The linear-polarization state of the CMB in 
a direction ${\mathbf {\hat n}}$ can be described by a symmetric
trace-free $2\times2$ tensor,
\begin{equation}
  {\cal P}_{ab}(\hatn)={1 \over 2} \left( \begin{array}{cc}
   \vphantom{1\over 2}Q(\hatn) & -U(\hatn) \sin\theta \\
   -U(\hatn)\sin\theta & -Q(\hatn)\sin^2\theta \\
   \end{array} \right),
\label{eq:whatPis}
\end{equation}
where the subscripts $ab$ are tensor indices, and $Q(\hatn)$ and
$U(\hatn)$ are the Stokes parameters.  Just as the temperature
map can be expanded in terms of spherical harmonics, the
polarization tensor can be expanded
\cite{KamKosSte97a,KamKosSte97b,SelZal97,ZalSel97},
\begin{equation}
      {{\cal P}_{ab}(\hatn)\over T_0} =
      \sum_{lm} \left[ a_{(lm)}^{{\rm G}}Y_{(lm)ab}^{{\rm
      G}}(\hatn) +a_{(lm)}^{{\rm C}}Y_{(lm)ab}^{{\rm C}}(\hatn)
      \right],
\label{eq:Pexpansion}
\end{equation}
in terms of tensor spherical harmonics, $Y_{(lm)ab}^{\rm G}$
and $Y_{(lm)ab}^{\rm C}$.  It is well known that a vector field
can be decomposed into a curl and a curl-free (gradient)
part.  Similarly, a $2\times2$ symmetric traceless
tensor field can be decomposed into a tensor
analogue of a curl and a gradient part; the $Y_{(lm)ab}^{\rm G}$
and $Y_{(lm)ab}^{\rm C}$ form a complete orthonormal basis for the
``gradient'' (i.e., curl-free) and ``curl'' components of the
tensor field, respectively.  
The mode amplitudes in Eq. (\ref{eq:Pexpansion}) are given by
\begin{eqnarray}
     a^{\rm G}_{(lm)}&=&{1\over T_0}\int d\hatn\,{\cal P}_{ab}(\hatn)\, 
                                         Y_{(lm)}^{{\rm G}
                                         \,ab\, *}(\hatn),\cr 
     a^{\rm C}_{(lm)}&=&{1\over T_0}\int d\hatn\,{\cal P}_{ab}(\hatn)\,
                                          Y_{(lm)}^{{\rm C} \,
                                          ab\, *}(\hatn), 
\label{eq:Amplitudes}
\end{eqnarray}
which can be derived from the orthonormality properties of these 
tensor harmonics \cite{KamKosSte97b}.
Thus, given a polarization map ${\cal P}_{ab}(\hatn)$, the G and 
C components can be isolated by first carrying out the
transformations in Eq. (\ref{eq:Amplitudes}) to the $a^{\rm
G}_{(lm)}$ and $a^{\rm C}_{(lm)}$, and then summing over the
first term on the right-hand side of Eq. (\ref{eq:Pexpansion})
to get the G component and over the second term to get the C
component. 
The two-point statistics of the combined
temperature/polarization (T/P) map are specified completely by
the six power spectra $C_\ell^{{\rm X}{\rm X}'}$
for ${\rm X},{\rm X}' = \{{\rm T,G,C}\}$, but parity invariance
demands that $C_\ell^{\rm TC}=C_\ell^{\rm GC}=0$.  Therefore,
the statistics of the CMB temperature-polarization map are
completely specified by the four sets of moments: $C_\ell^{\rm
TT}$, $C_\ell^{\rm TG}$, $C_\ell^{\rm GG}$, and $C_\ell^{\rm
CC}$.

Both density perturbations and gravitational waves will produce
a gradient component in the polarization.  However, to linear
order in small perturbations, only gravitational waves will
produce a curl component \cite{KamKosSte97a,SelZal97}.
The curl component thus provides a model-independent
probe of the gravitational-wave background.

In \scite{KamKos98} and \scite{JafKamWan00}, we studied the smallest IGW
amplitude that can be detected by CMB experiments parameterized
by a fraction of sky covered, the instrumental sensitivity
(parameterized by a noise-equivalent temperature $s$), and an
angular resolution.  We found that the sensitivity to IGWs was
maximized with a survey that covers roughly a
$5^\circ\times5^\circ$ patch of the sky (as indicated by the
solid curve in Fig. \ref{fig:limits}) and with an angular
resolution better than roughly $1^\circ$.  The smallest
detectable energy scale of inflation is then $E_{\rm infl} =
5\times10^{15}(s/25\, \mu{\rm K}~\sqrt{\rm sec})^{1/2} \, {\rm
GeV}$.  For reference, the instrumental sensitivity for MAP is
$O(100 \mu{\rm K}~\sqrt{\rm sec})$ and for the Planck satellite
$O(20 \, \mu{\rm K}~\sqrt{\rm sec})$.

However, since then, it has been pointed out that cosmic shear (CS),
gravitational lensing of the CMB due to large-scale structure
along the line of sight, can convert some of the curl-free
polarization pattern at the surface of last scatter into a curl
component, even in the absence of gravitational waves
\cite{ZalSel98}.  This
cosmic-shear--induced curl can thus be confused with that due to
gravitational waves.  In principle, the two can be distinguished
because of their different power spectra, as shown in
Fig. \ref{fig:powerspectrum}, but if the IGW amplitude is small,
then the separation becomes more difficult.  \scite{LewChaTur02},
\scite{KesCooKam02}, and \scite{KnoSon02} showed that when the
cosmic-shear confusion is taken into account, the smallest
detectable inflationary energy scale is $\simeq4\times10^{15}$
GeV.

The deflection angle due to
cosmic shear can in principle be mapped as a function of
position on the sky by studying higher-order correlations in the
measured CMB temperature and polarization
\cite{SelZal99,Hu01a,Hu01b,HuOka02,CooKes02}.  If
this deflection
angle is determined, then the polarization can be corrected and
the polarization pattern at the surface of last scatter can be
reconstructed.  \scite{KesCooKam02} and \scite{KnoSon02} found that
with such a reconstruction, the cosmic-shear--induced CMB curl
component can be reduced by roughly a factor of ten, as
indicated in Fig. \ref{fig:powerspectrum}.  This then leads to a
smallest inflationary energy scale that will produce a
detectable IGW signal in the CMB polarization curl.  The
conclusion is that the CMB-polarization signature of IGWs will
be undetectable, even with perfect detectors, if the energy
scale of inflation is smaller than $2\times10^{15}$ GeV.

Let us now suppose that this curl component was indeed
detected.  It would immediately tell us that the vacuum-energy
density during inflation was $(10^{15-16}\,{\rm GeV})^4$, and
thus that inflation probably had something to do with grand
unification.  However, there is possibly more that we
can learn.  Since the unifying high-energy physics responsible
for inflation presumably encompasses electroweak interactions as
a low-energy limit, and since the weak interactions are parity
violating, it is not unreasonable to wonder whether the physics
responsible for inflation is parity violating.
\scite{LueWanKam98} and \scite{Lep98} showed how
parity-violating observables could be constructed from a CMB
temperature-polarization map.  Moreover, examples were provided
of parity-violating terms in the inflaton Lagrangian that would
give rise to such signatures by, for example,
producing a preponderance of right- over left-handed
gravitational waves.

\section{Non-Gaussian initial conditions}

\begin{figure}[!t]
  \includegraphics[width=\columnwidth]{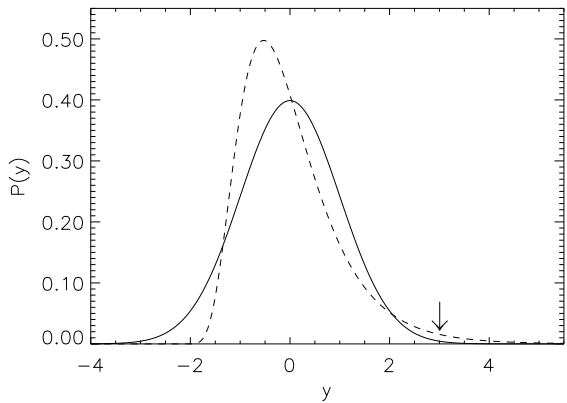}
  \caption{The solid curve shows a Gaussian distribution $P(y)$
     with unit variance, while the broken curve shows a
     non-Gaussian distribution with the same variance but 10
     times as many peaks with $y>3$.  This illustrates (a) how
     the cluster abundance can be dramatically enhanced with
     long non-Gaussian tails (since clusters form from rare
     peaks); and (b) that the dispersion of $y$ for $y>3$ is
     much larger for the non-Gaussian distribution than it is
     for the Gaussian distribution, and this will lead to a
     larger scatter in the formation redshifts and sizes of
     clusters of a given mass.  From \protect\scite{ourclusters}.}
  \label{fig:distributions}
\end{figure}

\begin{figure}[!t]
  \includegraphics[width=\columnwidth]{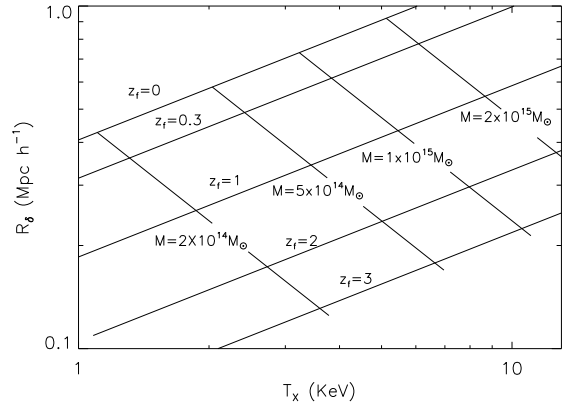}
  \caption{Mass and formation-redshift contours in the
size-temperature plane for $\Omega_0=0.3$ and $h=0.65$ obtained
{}from the spherical-top-hat model of gravitational collapse
discussed in the text.  It is clear from the figure that a
narrow (broad) spread in the formation redshift will yield a tight
(broad) size-temperature relation.  For larger $\Omega_0$, the $z_f=0$ contour 
remains the same, but the spacing between equi-$z_f$ contours
increases.  From \protect\scite{ourclusters}.}
  \label{fig:sizetemp}
\end{figure}

\begin{figure*}[!t]
  \includegraphics[width=2\columnwidth]{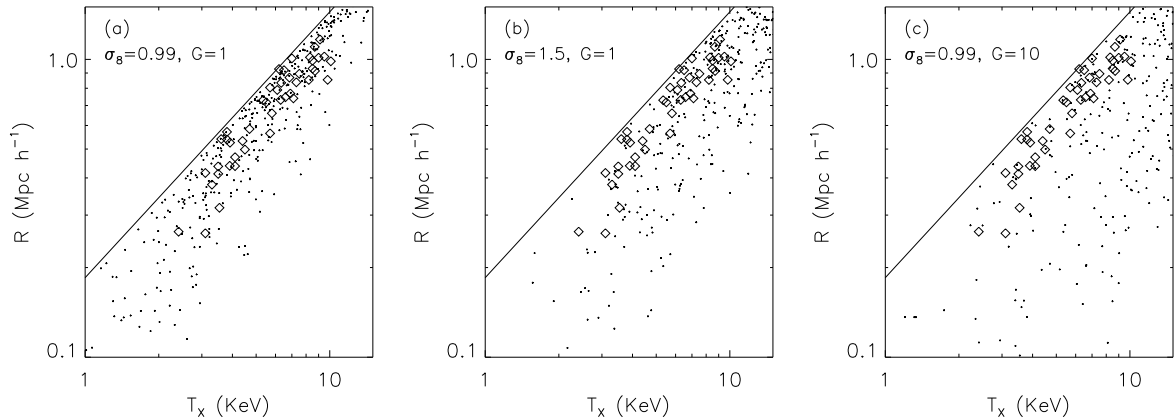}
  \caption{(a) Size-temperature distribution for LCDM and
     $\sigma_8=0.99$ and Gaussian initial conditions.  Each dot
     represents a simulated cluster, while the diamonds are
     data from M00.  The line shows the size-temperature
     relation expected for clusters today.  Panel (b) shows the same
     except that here we use $\sigma_8=1.5$.  Panel (c) shows the same
     as in (a) but with the skew-positive distribution of
     \protect\scite{RobGawSil00}.  From \protect\scite{ourclusters}.}
  \label{fig:basicresults}
\end{figure*}

The simplest single-scalar-field inflation models predict that
primordial perturbations have nearly Gaussian initial
conditions.  A small degree of non-Gaussianity generally arises
from self-coupling of the inflaton field, but this is expected
to be very tiny
\cite{SalBonBar89,Sal92,FalRanSre93,Ganetal94,Gan94}. More
complicated models of inflation, such as
two-field \cite{BarMatRio02}, warm \cite{Gupetal02}, or curvaton
\cite{Lytetal02} models may have small deviations from
perfectly Gaussian initial conditions, and higher-order
calculations of perturbation production suggest that
non-Gaussianity may be significant even in slow-roll models
\cite{Acqetal02}.  Although it is difficult, if not impossible,
to predict the exact amplitude and precise form of
non-Gaussianity from inflation, it is certainly reasonable to
search for it.

Perhaps the most intuitive place to look for primordial
non-Gaussianity is in the CMB.  Since the CMB temperature
fluctuations probe directly primordial density perturbations,
non-Gaussianity in the density field should lead to
proportionate non-Gaussianity in the temperature maps.  So, for
example, if the primordial distribution of perturbations is
skewed, then there should be a skewness in the temperature
distribution.  Alternatively, one can study the effects of
primordial non-Gaussianity in the distribution of mass in the
Universe today.  This should have the advantage that density
perturbations have undergone gravitational amplification and
should thus have a larger amplitude than in the early Universe.
However, we must keep in mind that the matter distribution today
is expected to be non-Gaussian, even if primordial perturbations
are Gaussian.  This can be seen just by noting that
gravitational infall can lead to regions---e.g., galaxies or
clusters---with densities of order 200 times the mean density,
while the smallest underdensity, a void, has a fractional
underdensity of only $-1$.  However, non-Gaussianity in the galaxy
distribution from gravitational infall from Gaussian initial
conditions can be calculated fairly reliably (for an excellent
recent review, see, e.g., \pcite{Roman}), and so the
distribution today can be checked for consistency with
primordial Gaussianity.  In \scite{verdeone}, we
studied the relative sensitivity of galaxy surveys and CMB
experiments to primordial non-Gaussianity.  We considered
two classes of non-Gaussianity:  in the first,
the gravitational potential is written $\phi({\mathbf
x},t)=g({\mathbf x},t) + \epsilon g^2({\mathbf x},t)$, where
$g(x)$ is a Gaussian random field (so $\phi$ becomes Gaussian as
$\epsilon \rightarrow 0$); such a form of non-Gaussianity arises
in some inflation models.  In the second, the
fractional density perturbation is written  $\delta({\mathbf
x},t)=g({\mathbf x},t) + \epsilon g^2({\mathbf x},t)$; this
approximates the form of non-Gaussianity expected from
topological defects.  We then
determined what the smallest detectable $\epsilon$ would be for
both cases for future galaxy surveys and for CMB experiments.  We
found that in both cases the CMB would provide a more sensitive
probe of $\epsilon$.  Conversely, if the CMB turns out to be
consistent with primordial Gaussianity, then for all practical
purposes, the galaxy distribution can safely be assumed to arise
from Gaussian initial conditions.

Experimentally, the bispectrum from 2dF \cite{Veretal02} and the
Sloan Digital Sky Survey \cite{Szaetal02} have now been studied
and found to be consistent with Gaussian initial conditions.  A
tentative claim of non-Gaussianity in the the COBE-{\sl DMR}
maps \cite{Feretal98} in great excess of slow-roll-inflationary
expectations \cite{KamWan00,GanMar00} was later found to be due
to a very unusual and subtle systematic effect in the data
\cite{Banetal00}.

The abundances of clusters provide other avenues
toward detecting primordial non-Gaussianity.  Galaxy clusters,
the most massive gravitationally bound objects in the Universe
presumably form at the highest-density peaks in the primordial
density field, as indicated schematically in
Fig. \ref{fig:distributions}.  Now suppose that instead of a
Gaussian primordial distribution, we had a distribution with
positive skewness, as shown in Fig. \ref{fig:distributions}.  In
this case, we would expect there to be more high-density peaks,
even for a distribution with the same variance, and thus more
clusters \cite{RobGawSil00}.  Thus, the cluster abundance can be
used to probe this type of primordial non-Gaussianity.

Just as clusters are rare objects in the Universe today,
galaxies were rare at redshifts $z\gtrsim3$.  In
\cite{Veretal01a}, we considered the use of abundances of
high-redshift galaxies as probes of primordial
non-Gaussianity. We also found an expression that relates the
excess abundance of rare objects to the $\epsilon$ parameter in
the models discussed above and were thus able to compare the
sensitivities of cluster and high-redshift-galaxy counts and the
CMB to non-Gaussianity in the models considered above.  We found
that although the CMB was expected to be superior in detecting
non-Gaussianity in the gravitational potential, the
high-redshift-galaxy abundances may do better with
non-Gaussianity in the density field.

In addition to producing more clusters, such a skew-positive
distribution might change the distribution of the properties of
such objects.  In \scite{ourclusters}, we considered the
size-temperature relation.  If we model the formation of a
cluster as a spherical top-hat collapse, then the virial radius
and the virial temperature can be determined as a function of
the halo mass and the collapse redshift.  We then modeled the
x-ray-emitting gas to relate its size and temperature to the
virial radius and temperature of the halo in which it 
lives in order to obtain better estimates for the x-ray isophotal radii
and x-ray temperatures that are measured.
Fig. \ref{fig:sizetemp} shows resulting contours of constant
mass and constant collapse redshift in the cluster
size-temperature plane.  As shown there, halos that collapse
earlier should lead to hotter and smaller clusters, and more
massive halos should be hotter and bigger.  If the primordial
distribution is skew-positive, then the halos that house
clusters will collapse over a wider range of redshifts, and as
indicated in Fig. \ref{fig:basicresults}, this will lead to a
broader scatter than in the size-temperature relation than is
observed.  In this way, primordial distributions with a large
positive skewness can be constrained.

\section{Broken Scale Invariance and Galactic Substructure}

\begin{figure}[!t]
  \includegraphics[width=\columnwidth]{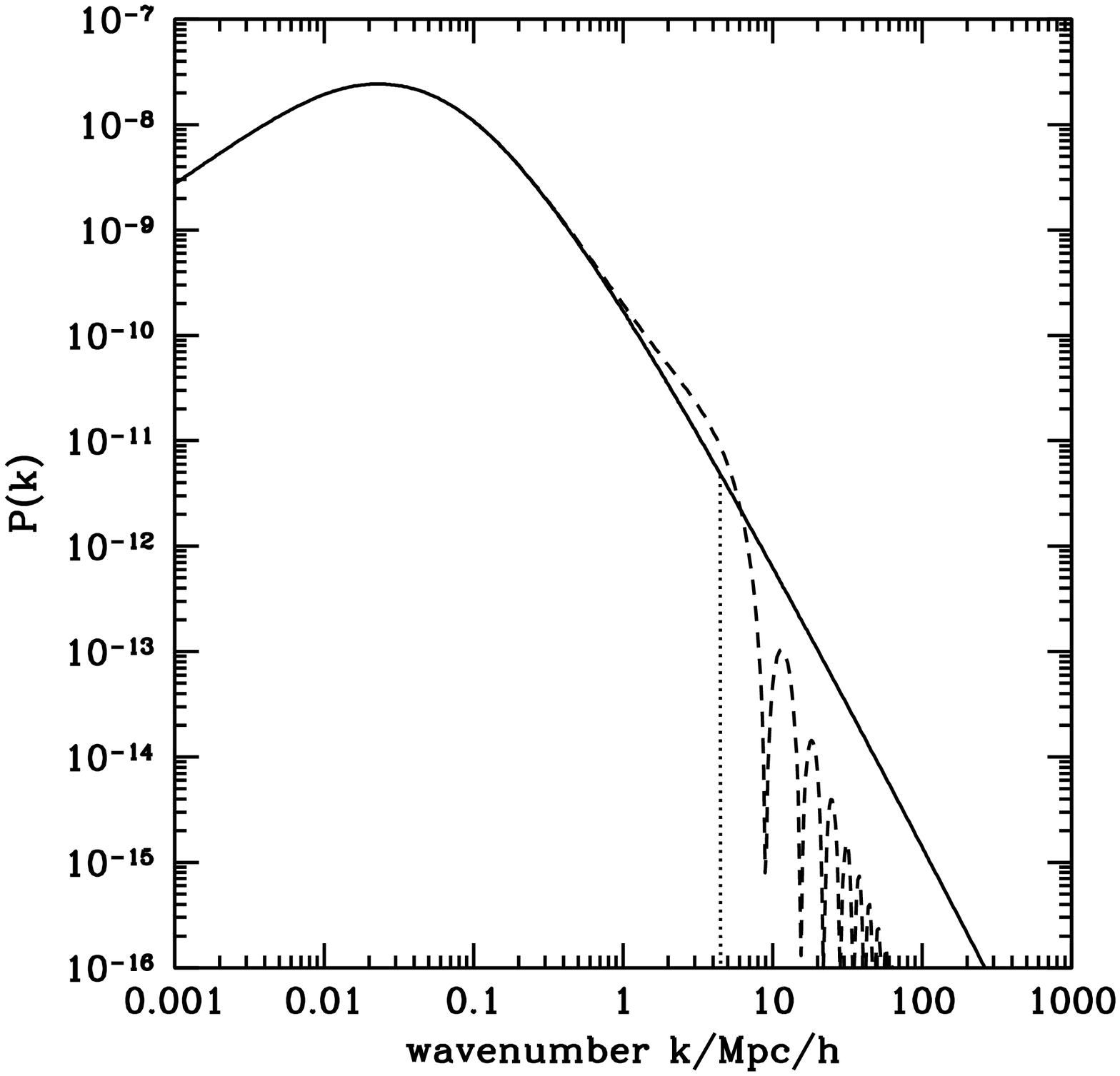}
  \includegraphics[width=\columnwidth]{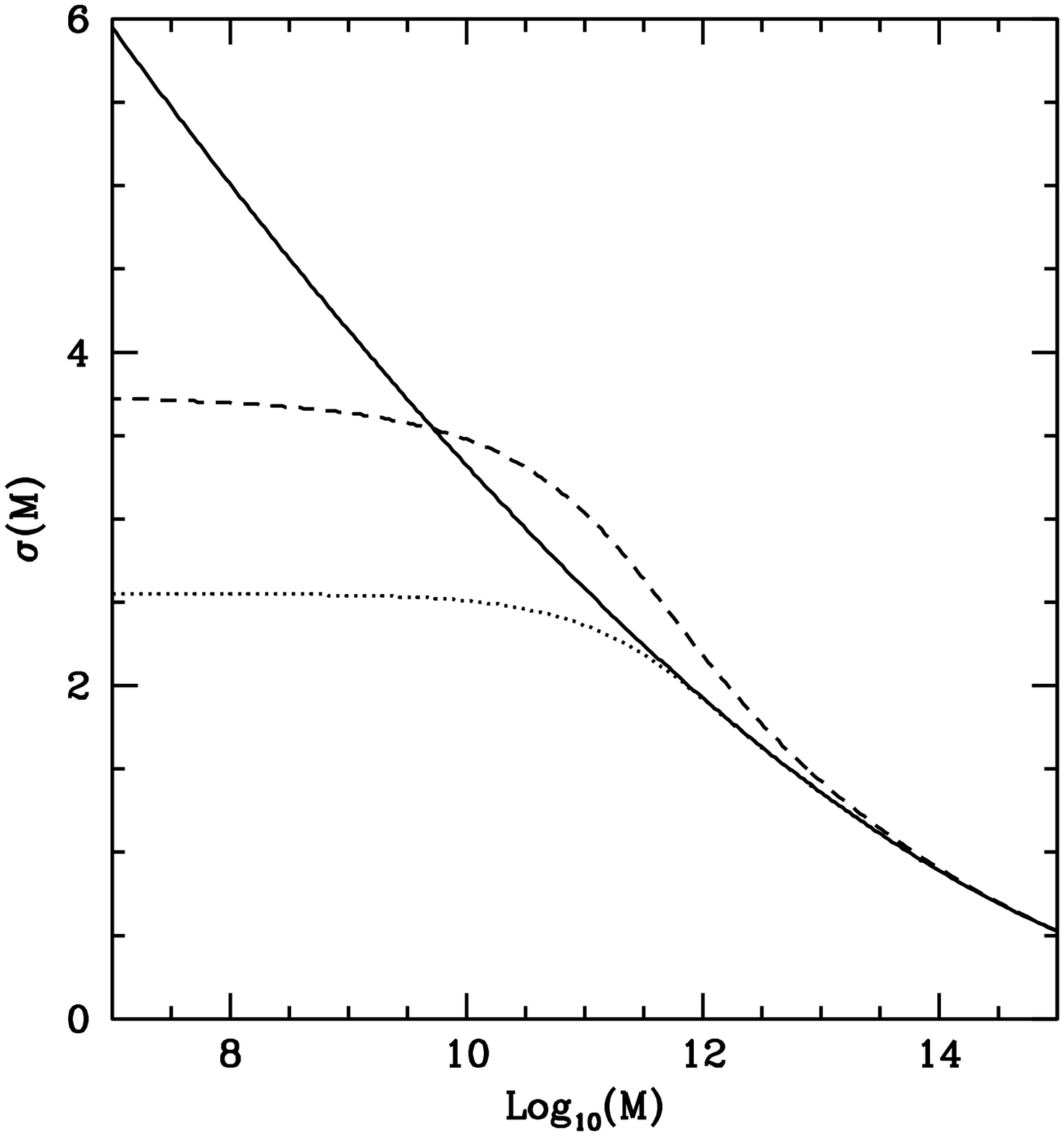}
  \caption{The upper panel shows the power spectrum for
  an LCDM model (solid curve), for a model in which the
  power spectrum is arbitrarily cut off at $k=4.5\,h$ Mpc$^{-1}$
  (dotted curve), and the broken-scale-invariance inflation
  model (dashed curve).  The lower panel shows the rms mass
  fluctuation as a function of the enclosed mean mass $M$ for
  these three models.  From \protect\scite{KamLid00}.}
  \label{fig:power}
\end{figure}

\begin{figure}[!t]
  \includegraphics[width=\columnwidth]{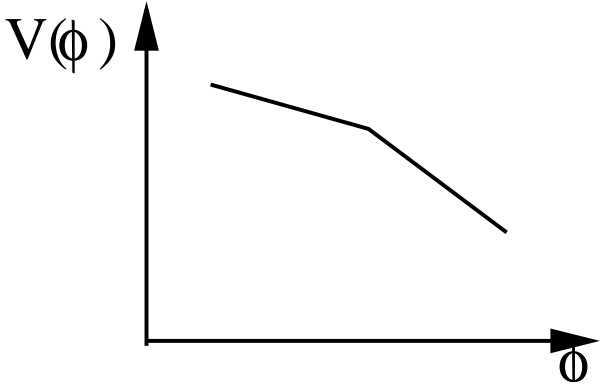}
  \caption{An inflaton potential with a break in the first derivative.}
  \label{fig:break}
\end{figure}

\begin{figure}[!t]
  \includegraphics[width=\columnwidth]{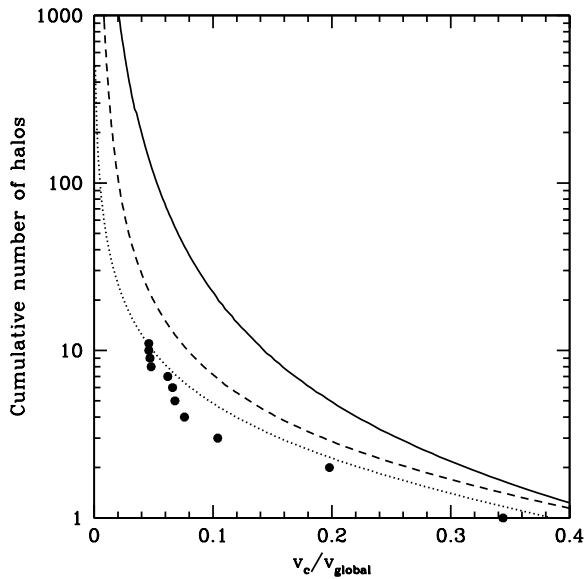}
  \caption{The cumulative number of mini-halos for the power spectra
     shown in Fig.~\protect\ref{fig:powerspectrum} as a function 
     of the circular speed $v_{{\rm c}}$ of the halo divided by the
     circular speed $v_{\rm global}$ of the Galactic halo.  The
     points show the Milky Way satellites.  From
     \protect\scite{KamLid00}, after \protect\scite{Mooetal99}.}
  \label{fig:N}
\end{figure}

The final probe of inflation that we will discuss here is
motivated by the galactic-substructure problem.  N-body
simulations of structure formation with a standard
inflation-inspired scale-free spectrum of primordial
perturbations predict far more substructure, in the form of
dwarf galaxies, in galactic halos than is observed in the
Milky Way halo, as indicated in Fig. \ref{fig:N}
\cite{KlyKraVal99,Mooetal99}.  Although a number
of possible astrophysical mechanisms for suppressing this
small-scale power have been proposed
(e.g., \pcite{BulKraWei00,Benetal02,Stoetal02}),
there is still no general consensus on whether they are
sufficiently effective to eliminate the problem.

Another possible explanation of the observed dearth of dwarf
galaxies is a small-scale suppression of power that could occur
if the inflaton potential has a sharp feature, like that shown
in Fig. \ref{fig:break} \cite{KamLid00,Yok00}.
According to inflation, primordial density perturbations are
produced by quantum fluctuations in the inflaton, the scalar
field responsible for inflation.  Moreover, the details of the
power spectrum $P(k)$ of density perturbations (as shown in Fig.
\ref{fig:power}) is determined by the shape
$V(\phi)$ of the inflaton potential.  The amplitude of a given
Fourier mode of the density field is proportional to the value
of $V^{3/2}/V'$, where $V'$ is the first derivative of the
inflaton potential, at the time that the perturbation exited the
horizon.  In most models, the inflaton potential is smooth and
this leads to a power spectrum of perturbations that is very
nearly a power law---Ref. \cite{reconstruction} explains very
nicely how the amplitude and slope of the inflaton potential can
be reconstructed in this case.

However, suppose that for some reason there is a break in the
inflaton potential, as shown in Fig. \ref{fig:break}, and the
slope increases suddenly as the inflaton roles down the
potential.  In this case, $V'$ increases suddenly, and
since the density-perturbation amplitude is $\propto 1/V'$, the
density-perturbation amplitude on small scales (those that exit
the horizon last) will be suppressed, as indicated by the dashed
curve in Fig. \ref{fig:power}.  The wiggles in the dashed curve
are ringing in Fourier space that results from the sharpness of
the feature.  If it is smoothed out, then a power spectrum more
like the dotted curve in Fig. \ref{fig:power} becomes possible.

With the three power spectra in the upper panel of
Fig. \ref{fig:power}, the rms mass fluctuation $\sigma(M)$ on a
mass scale $M$ can be calculated, as shown in the lower panel of
Fig. \ref{fig:power}.  With the scale-invariant spectrum,
$\sigma(M)$ keeps rising as we go to smaller and smaller masses,
leading to substructure on smaller scales.  However, if power is
suppressed on small scales, then $\sigma(M)$ ceases to rise (or
rises only very slowly) at small $M$ implying the absence (or
suppression) of halos of these small masses.

Given $\sigma(M)$ for these three power spectra, the abundance
of sub-halos in a typical galaxy-mass halo of $10^{12}\,
M_\odot$ can be calculated with the extended Press-Schechter
formalism.  Results of this calculation are shown in
Fig. \ref{fig:N}.  As a check, the approximation reproduces well
the numerical-simulation results for the scale-free spectrum.
For the power spectra with broken scale invariance, the
abundance of low-mass substructure is reduced and brought into
reasonable agreement with the observed ten or so Milky Way
satellites, without violating consistency with constraints from
the Lyman-alpha forest \cite{KamLid00,WhiCro00}.

Is such a break to be expected theoretically?  Probably not, and
there are probably simpler explanations for the shortfall that
involve more conventional astrophysics.  Still, these
calculations show that by studying and understanding
galactic substructure, we learn about the shape of the inflaton
potential toward the end of inflation in a way that complements
the information from earlier epochs of inflation that comes from
larger scales.

\section{Comments}

In this talk, I have briefly reviewed several avenues for
probing some of the non-standard predictions of inflation.
There are yet other tests---in particular, searches for a small
entropy component of primordial perturbations---that I have
not had time to discuss.
Since our current models of inflation, as successful as they may
be in terms of the flatness and primordial perturbations, are
really no more than toy models, any realistic model of inflation
must be, in some sense, non-standard.  Although I believe that
forthcoming CMB satellite experiments are likely to further
confirm a flat Universe, small deviations from the toy model must
sooner or later arise.  Unfortunately, no single inflation model
is sufficiently well motivated to predict with any certainty the
existence or detectability of any of these signatures.  However,
as we press forward with the study of galactic structure and
formation, large-scale structure, and the CMB, we will
inevitably develop the capabilities for learning more about
inflation.

\medskip
I thank my collaborators on the work reported here, and I
acknowledge the hospitality of the Kavli Institute for
Theoretical Physics.  This work was supported at Caltech by NSF
AST-0096023, NASA NAG5-9821, and DoE DE-FG03-92-ER40701, and at
the KITP by NSF PHY99-07949.


\begin{thebibliography}

\bibitem[Abbott \& Wise<1984>]{AbbWis84} Abbott, L. F. \& Wise,
     M. 1984, Nucl. Phys., B244, 541

\bibitem[Acquaviva et al.{}<2002>]{Acqetal02} Acquaviva, V., et al. 2002,
     astro-ph/0209156 

\bibitem[Albrecht \& Steinhardt<1982>]{AlbSte82} Albrecht, A. \&
     Steinhardt, P. J. 1982, Phys. Rev. Lett., 48, 1220

\bibitem[Banday et al.{}<2000>]{Banetal00} Banday, A. J., Zaroubi, S.,
     G\'orski, K. M. 2000, ApJ, 533, 575

\bibitem[Bardeen et al.{}<1983>]{BarSteTur83} Bardeen, J. M,
     Steinhardt, P. J, \& Turner M. S. 1983, Phys.~Rev.~D, 46, 645

\bibitem[Bartolo et al.{}<2002>]{BarMatRio02} Bartolo, N., Matarrese, S., \&
     Riotto, A. 2002, Phys. Rev. D, 65, 103505

\bibitem[Benson et al.{}<2002>]{Benetal02} Benson, A. J., et al. 2002,
     MNRAS, 333, 177

\bibitem[Bernardeau et al.{}<2002>]{Roman} Bernardeau, F., et
     al. 2002, Phys. Rep., 367, 1

\bibitem[de Bernardis et al.{}<2000>]{deBetal00} de Bernardis,
     P., et al. 2002, Nature, 404, 955

\bibitem[Bullock et al.{}<2000>]{BulKraWei00} Bullock, J. S.,
     Kravtsov, A. V., \& Weinberg, D. H. 2000, ApJ, 539, 517

\bibitem[Cooray \& Kesden<2002>]{CooKes02} Cooray, A. \& Kesden,
     M., 2002, astro-ph/0204068

\bibitem[Falk et al.{}<1993>]{FalRanSre93} Falk, T., Rangarajan,
     R., \& Srednicki, M. 1993, ApJ, 403, L1

\bibitem[Ferreira et al.{}<1998>]{Feretal98} Ferreira, P. G., Magueijo, J.,
     \& G\'orski, K. M. 1998, ApJ, 503, L1

\bibitem[Gangui et al.{}<1994>]{Ganetal94} Gangui, A., et
     al. 1994, ApJ, 430, 447

\bibitem[Gangui<1994>]{Gan94} Gangui, A. 1994, Phys.~Rev.~D, 50, 3684

\bibitem[Gangui \& Martin<2000>]{GanMar00} Gangui, A. \& Martin, J. 2000,
     MNRAS, 313, 323

\bibitem[Gupta et al.{}<2002>]{Gupetal02} Gupta, S., et al. 2002,
     Phys. Rev. D, 66, 043510

\bibitem[Guth<1981>]{Gut81} Guth, A. H. 1981, Phys.~Rev.~D, 28, 347

\bibitem[Guth \& Pi<1982>]{GutPi82} Guth, A. H \& Pi,
     S.-Y. 1982, Phys. Rev. Lett., 49, 1110

\bibitem[Halverson et al.{}<2002>]{Haletal02} Halverson, N. W., et
     al. 2002, ApJ, 568, 38

\bibitem[Hanany et al.{}<2000>]{Hanetal00} Hanany, S., et
     al. 2000, ApJ, 545, L5

\bibitem[Hawking<1982>]{Haw82} Hawking, S. W. 1982, Phys. Lett.,
     B115, 29

\bibitem[Hu<2001a>]{Hu01a} Hu, W. 2001a, Phys. Rev. D, 64, 083005

\bibitem[Hu<2001b>]{Hu01b} Hu, W. 2001b, ApJ, 557, L79
	
\bibitem[Hu \& Okamoto<2002>]{HuOka02} Hu, W. \& Okamoto, T. 2002,
     ApJ, 574, 566

\bibitem[Jaffe et al.{}<2000>]{JafKamWan00} Jaffe, A. H.,
     Kamionkowski, M., \& Wang, L. 2000, Phys. Rev. D, 61, 083501

\bibitem[Kamionkowski et al.{}<1997a>]{KamKosSte97a}
     Kamionkowski, M., Kosowsky, A., \& Stebbins, A. 1997a,
     Phys. Rev. Lett., 78, 2058. 

\bibitem[Kamionkowski et al.{}<1997b>]{KamKosSte97b}
     Kamionkowski, M., Kosowsky, A., \& Stebbins, A. 1997b,
     Phys. Rev. D, 55, 7368

\adjustfinalcols

\bibitem[Kamionkowski \& Kosowsky<1998>]{KamKos98} Kamionkowski,
     M. \& Kosowsky, A. 1998, Phys.~Rev.~D, 67, 685

\bibitem[Kamionkowski \& Kosowsky<1999>]{KamKos99} Kamionkowski,
     M. \& Kosowsky, A. 1999, Ann. Rev. Nucl. Part. Sci., 49, 77

\bibitem[Kamionkowski \& Liddle<2000>]{KamLid00} Kamionkowski,
     M. \& Liddle, A. R., 2000, Phys. Rev. Lett., 84, 4525

\bibitem[Kesden et al.{}<2002>]{KesCooKam02} Kesden, M., Cooray, A., \&
     Kamionkowski, M. 2002, Phys. Rev. Lett., 89, 011304

\bibitem[Klypin et al.{}<1999>]{KlyKraVal99} Klypin, A. A.,
     Kravtsov, A. V., \& Valenzuela, O. 1999, ApJ, 522, 82

\bibitem[Knox \& Song<2002>]{KnoSon02} Knox, L. \& Song, Y.-S. 2002,
     Phys. Rev. Lett., 89, 011303

\bibitem[Lepora<1998>]{Lep98} Lepora, N., qr-qc/9812077

\bibitem[Lewis et al.<2002>]{LewChaTur02} Lewis, A., Challinor,
     A., \& Turok, N. 2002, Phys. Rev. D, 65, 023505

\bibitem[Linde<1982a>]{Lin82a} Linde, A. D. 1982a, Phys. Lett.,
     B108 389

\bibitem[Linde<1982b>]{Lin82b} Linde, A. D. 1982b, Phys. Lett., B116, 335

\bibitem[Lidsey et al.{}<1997>]{reconstruction} Lidsey, J. E.,
     et al. 1997, Rev. Mod. Phys., 69, 373

\bibitem[Lue et al.{}<1998>]{LueWanKam98} Lue, A., Wang, L., \& Kamionkowski,
     M. 1998, Phys. Rev. Lett., 83, 1503

\bibitem[Lyth et al.{}<2002>]{Lytetal02} Lyth, D., Ungarelli,
     C., \& Wands, D. 2002, astro-ph/0208055

\bibitem[Mason et al.{}<2002>]{Masetal02} Mason, B. S., et al.,
     astro-ph/0205384 

\bibitem[Miller et al.{}<1999>]{Miletal99} Miller, A. D., et
     al. 1999, ApJ, 524, L1

\bibitem[Moore et al.{}<1999>]{Mooetal99} Moore, B., et
     al. 1999, ApJ, 524, L19 

\bibitem[Robinson et al.{}<2000>]{RobGawSil00} Robinson, J., Gawiser, E., \&
     Silk, J. 2002, ApJ, 2000, 532, 1

\bibitem[Salopek et al.{}<1989>]{SalBonBar89} Salopek, D. S.,
     Bond, J. R., \& Bardeen, J. M. 1989, Phys.~Rev.~D, 40, 1753

\bibitem[Salopek<1992>]{Sal92} Salopek, D. S. 1992, Phys.~Rev.~D, 45, 1139

\bibitem[Seljak \& Zaldarriaga<1997>]{SelZal97} Seljak, U. \&
     Zaldarriaga, M. 1997, Phys. Rev. Lett., 78, 2054

\bibitem[Seljak \& Zaldarriaga<1999>]{SelZal99} Seljak, U. \&
     Zaldarriaga, M. 1999, Phys. Rev. Lett., 82, 2636

\bibitem[Starobinsky<1982>]{Sta82} Starobinsky, A. A. 1982,
     Phys. Lett., B117, 175

\bibitem[Stoehr et al.{}<2002>]{Stoetal02} Stoehr, F., et
     al. 2002, astro-ph/0203342.

\bibitem[Szapudi et al.{}<2002>]{Szaetal02} Szapudi, I., et al. 2002, ApJ,
     570, 75

\bibitem[Verde et al.{}<2000>]{verdeone} Verde, L., et al. 2000, MNRAS,
     313, L141

\bibitem[Verde et al.{}<2001a>]{Veretal01a} Verde, L., et
     al. 2001, MNRAS, 325, 412

\bibitem[Verde et al.{}<2001b>]{ourclusters} Verde, L., et
     al. 2001, MNRAS, 321, L7

\bibitem[Verde et al.{}<2002>]{Veretal02} Verde, L., et al. 2002,
    MNRAS, 335, 432

\bibitem[Wang \& Kamionkowski<2000>]{KamWan00} Wang, L. \&
     Kamionkowski, M. 2000, Phys. Rev. D, 61, 063504

\bibitem[White \& Croft<2000>]{WhiCro00} White, M. \& Croft,
     R. A. C. 2000, ApJ, 539, 497

\bibitem[Yokoyama<2000>]{Yok00} Yokoyama, J. 2000, Phys. Rev. D, 62, 123509

\bibitem[Zaldarriaga \& Seljak<1997>]{ZalSel97} Zaldarriaga,
     M. \& Seljak, U. 1997, Phys.~Rev.~D, 55, 1830

\bibitem[Zaldarriaga \& Seljak<1998>]{ZalSel98} Zaldarriaga,
     M. \& Seljak, U. 1998, Phys. Rev. D, 58, 023003





\end{thebibliography}
\end{document}